# Generation of light with controllable spatial patterns via the sum frequency in quasi-phase matching crystals

Zhi-Yuan Zhou[1,2,#1], Yan Li[1,2,#1], Dong-Sheng Ding[1,2], Yun-Kun Jiang[3], Wei Zhang[1,2], Shuai Shi[1,2], Bao-Sen Shi[1,2,*] and Guang-Can Guo[1,2]

[1]*Key Laboratory of Quantum Information, University of Science and Technology of China, Hefei, Anhui 230026, China*

[2]*Synergetic Innovation Center of Quantum Information & Quantum Physics, University of Science and Technology of China, Hefei, Anhui 230026, China*

[3]*College of Physics and Information Engineering, Fuzhou University, Fuzhou 350002, China*

[*]*Corresponding author: drshi@ustc.edu.cn*

Light beams with extraordinary spatial structures, such as the Airy beam (AB), the Bessel-Gaussian beam (BGB) and the Laguerre-Gaussian beam (LGB), are widely studied and applied in many optical scenarios. We report on preparation of light beams with controllable spatial structures through sum frequency generation (SFG) using two Gaussian pump beams in a quasi-phase matching (QPM) crystal. The spatial structures, including multi-ring-like BGB, donut-like LGB, and super-Gaussian-like beams, can be controlled periodically via crystal phase mismatching by tuning the pump frequency or crystal temperature. This phenomenon has not been reported or discussed previously. Additionally, we present numerical simulations of the phenomenon, which agree very well with the experimental observations. Our findings give further insight into the SFG process in QPM crystals, provide a new way to generate light with unusual spatial structures, and may find applications in the fields of laser optics, all-optical switching, and optical manipulation and trapping.

There are many possible analytical solutions to the Maxwell equation in the paraxial approximation, including the Airy beam [1], the Bessel-Gaussian beam [2], the Hermite-Gaussian beam [3] and the Laguerre-Gaussian beam [4]. These solutions formed an orthogonal and closed basis for a paraxial propagating light beam. The peculiar properties of these beams shown in their propagation, diffraction and interference have been widely studied [5–10]. The extraordinary spatial structures of these beams make them suitable for applications in optical manipulation and trapping [11, 12], high speed and high capacity optical communications [13, 14], high precision optical measurements [15], and quantum information processing [16–20]. Sum frequency generation (SFG) is a method that is broadly used to

---

[#1] These two authors have contributed equally to this article.

extend the available wavelength range of light, with applications that include generation of new lasers in special wavelength regions [21–23], up-conversion detection of single photons [24–26] and up-conversion-based optical sensing [27]. Two types of nonlinear crystal are usually used for SFG: birefringent phase-matched crystals and quasi-phase matched (QPM) crystals. QPM crystals have the advantages of high effective nonlinear coefficients and no walk-off effect, which make them most suitable for high efficiency SFG.

By pumping a periodically poled potassium titanyl phosphate (PPKTP) crystal with two Gaussian beams, we can generate light beams with different spatial structures, such as the multi-ring-like BGB, the donut-like LGB, and a super-Gaussian-like beam through SFG. The spatial structures of these beams are determined by the phase mismatching of the crystal and the focusing parameters of the two pump beams. More interestingly, the spatial structure of the SFG beam varies periodically with changes in the phase mismatch, which is controlled by tuning of the frequencies of the pump lasers or the crystal temperature. Additionally, our detailed numerical simulations clearly show that this phenomenon depends on the focusing parameters of the pump beams. In addition, we have also observed asymmetrical behavior in SFG light in its spatial structure between the positive and negative phase mismatch regions. We have also simulated the SFG process numerically, and the simulation results match the results of the experimental observations very well. The experimental details and simulation results are described in the sections that follow.

The experimental scheme is illustrated in Figure 1. The wavelengths of the two pump beams are 1550 nm and 795 nm. The 1550 nm beam is from a diode laser (Toptica prodesign) and is amplified by an erbium-doped fiber amplifier (EDFA). The 795 nm laser beam is from a Ti: sapphire laser (MBR110, Coherent). The two beams are focused collinearly into a PPKTP crystal by lenses L1 and L2. The PPKTP crystal has dimensions of 1 mm×2 mm×10 mm, and is designed for SFG of 795 nm and 1550 nm to generate a beam at 525.5 nm, with a periodic poling period of 9.375 μm. The SFG light is collimated using lens L3 and is detected with a commercial charge-coupled device (CCD) camera after the pump beams are filtered out. The crystal temperature is controlled by a homemade temperature controller with stability of ±2 mK.

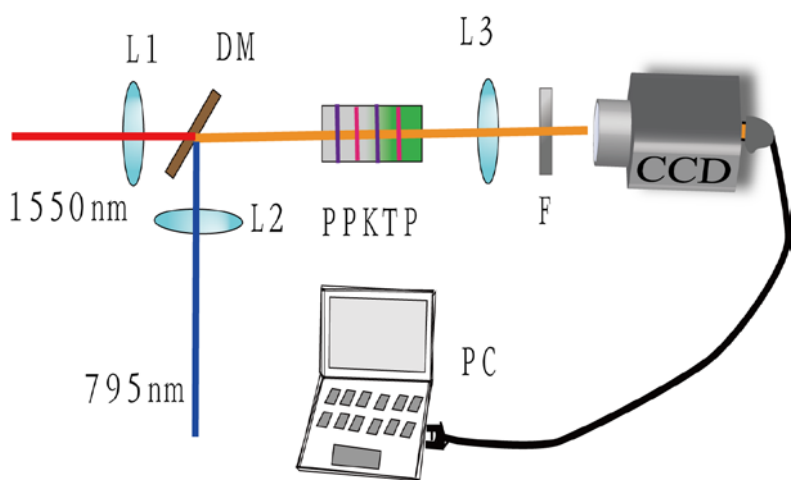

Figure 1. Simplified experimental setup. L1-L3: lenses; DM: dichromatic mirror; PPKTP: periodically poled KTP; F: filter; CCD: charge coupled device camera; PC: personal computer.

We first provide a theoretical description of SFG. To investigate SFG in QPM crystals, coupled wave

functions are used to describe the interaction of the waves as follows [28, 29]:

$$\begin{cases} \dfrac{dE_{1z}}{dx} = iK_1 E_{2z}^* E_{3z} e^{-i\Delta kx} \\ \dfrac{dE_{2z}}{dx} = iK_2 E_{1z}^* E_{3z} e^{-i\Delta kx} \\ \dfrac{dE_{3z}}{dx} = iK_3 E_{1z} E_{2z} e^{i\Delta kx} \end{cases}, \quad (1)$$

where $E_{jz}(j=1,2,3)$ represents the amplitudes of the three waves involved and $z$ denotes the polarization; the $K$-coefficients are $K_j = \dfrac{2\omega_j d_{33}}{\pi n_j c}(j=1,2,3)$ where $\omega_j$ is the angular frequency, $n$ is the refractive index, $c$ is the speed of light, and $d_{33}$ is the nonlinear coefficient of KTP; $\Delta k = k_{3z} - k_{1z} - k_{2z} - G_m$, where $k$ represents the wave vector. Under phase matching conditions, $\Delta k = 0$, which means that the momentum mismatch is fully compensated by the reciprocal vector $G_m$ of the QPM crystals. For Gaussian pump beams, the beam amplitudes $E_{jz}(j=1,2)$ can be expressed as [30]

$$E_{jz}(r,x) = \dfrac{A_j}{1-ix/X_{0j}} \exp\left(-\dfrac{r^2/\omega_{0j}^2}{1-ix/X_{0j}}\right) \quad (j=1,2) \quad (2)$$

where $A_j$ is a constant, $x$ is the propagation direction, $r = \sqrt{y^2 + z^2}$ represents the transverse coordinate, $\omega_{0j}$ denotes the beam waist and $X_{0j} = \pi \omega_{0j}^2 / \lambda_j$ is the Rayleigh range, where $\lambda_j$ is the wavelength.

Given the focusing parameters of the pump beams and the phase mismatch, the spatial structures of the SFG beam can be simulated numerically using a combination of equations (1) and (2). The experimental and simulated results are shown in Figure 2. The first and second rows are the experimental results. The first row shows spatial structures obtained by the CCD by tuning the crystal temperature towards its phase matching temperature. These five images represent a typical change period for the spatial structure. When we change the crystal temperature gradually, an outer ring first appears in the SFG light (like a BGB; see columns a and b), and then the outer ring becomes brighter while the central region becomes darker with the changing temperature. When the crystal temperature is set to a specific value, the central region then vanishes, and a donut structure (see column c) is observed. After that, the central intensity of the SFM light is then gradually retrieved by further changes in the temperature of the crystal, and the intensity distribution becomes flat at a certain temperature, as per a super-Gaussian beam (see column d). Finally, the beam intensity distribution returns to a Gaussian shape (see column e) again. The five images shown in the second row are the intensity distributions in the horizontal direction across the centers of the images. The third and fourth rows are the results of the numerical simulations corresponding to the first and second rows, respectively, based on our experimental parameters. The two sets of results show good agreement with each other. The differences between the colors of the first and third rows come from a constant multiplier that exists between the experiments and the numerical simulations of the processes. For a detailed overview of the processes of this phenomenon, readers can refer to the video included in the supplementary information. The above phenomenon can also be observed by tuning of the wavelength of either pump beam, which has the same function as tuning of the crystal temperature; the phase mismatching can be changed using either of the two methods. The results obtained by tuning of the pump wavelengths are therefore not shown in the text.

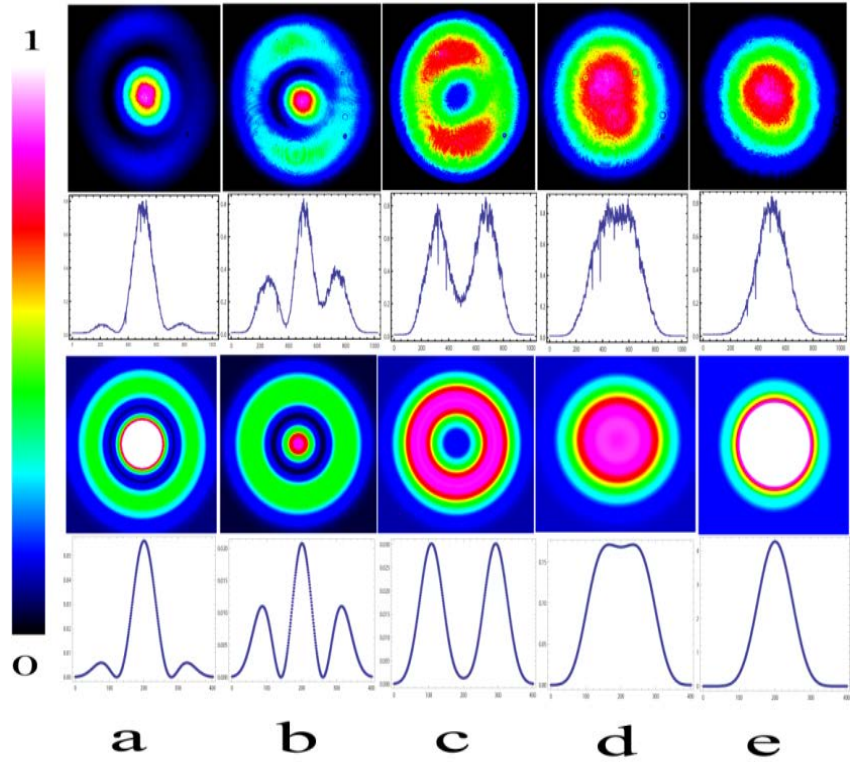

Figure 2. Experimental and simulation results. The first row shows the images obtained at different temperatures (46.2, 46.9, 47.6, 48.5, and 50.0°C for images a, b, c, d and e, respectively). The second row shows the corresponding intensity distributions in the horizontal direction across the centers of the images. The third and fourth rows are the simulation results corresponding to the experimental results of the first and second rows, respectively. In the simulations, the pump beam waists are 70 μm (795 nm) and 40 μm (1550 nm), and the phase mismatches represented by $\Delta k$ are (5.00, 4.65, 4.10, 3.20, 0.00)×$10^{-4}$ /μm for a, b, c, d, and e, respectively, in the third and fourth rows.

The phenomena observed above depend on the focusing parameters of the pump beams. The total phase in the third formula in equation (1) is $\Delta\varphi = \alpha_1 + \alpha_2 + \Delta k x$ ; here, $\alpha_j = \tan^{-1}(x/X_{0j}) - xr^2/X_{0j}\omega_{0j}^2[1+(x/X_{0j})^2]$ (j=1, 2) describe the phase parts of the two pump beams. $\alpha_j$ are spatially-dependent phases and relate to the beam waists of the pump beams.

For different focusing parameters, the same phenomena as that shown in Figure 2 could be observed. The only obvious difference is a shift in the phase mismatch. The effects of the focusing parameters were simulated numerically, and the results are shown in Figure 3, where the intensity distributions across the center of the SFG beam are shown. In the simulation, the 1550 nm pump beam waist is kept to a constant value of 40 μm, and the phase mismatch parameter $\Delta k$ is unchanged at 4.10×$10^{-4}$ /μm; the 795 nm pump beam waist changes from 30 μm to 80 μm with steps of 10 μm, and the results correspond to Figure 3a to f. The spatial phase depends on the focusing parameters: for tight focusing, the phase gradient is greater than that for weak focusing, inducing an intense change in the intensity profile, as shown in Figure 3a to d. The intensity profiles vary little under weak focusing conditions, as shown in Figure 3e and f.

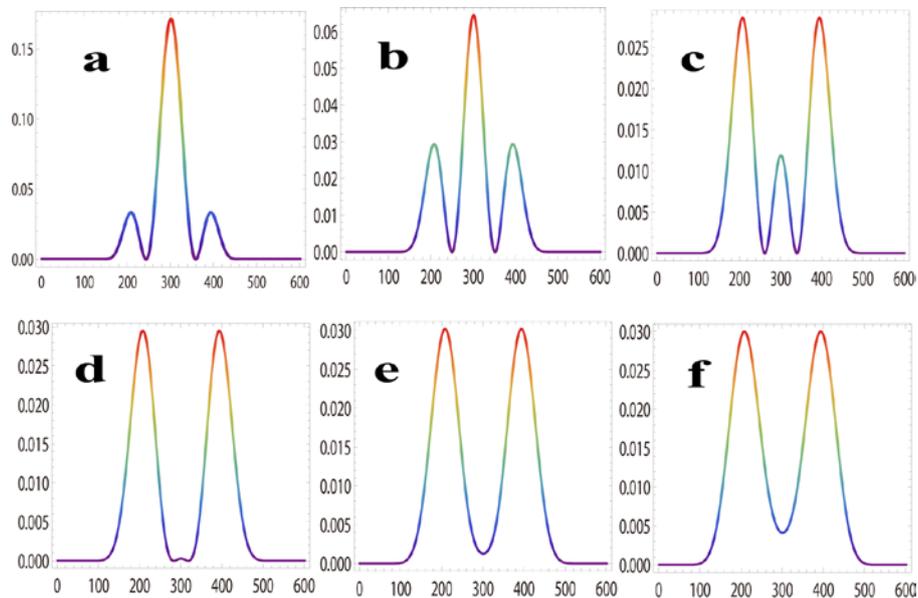

Figure 3. Simulation results of the dependence on the focusing parameters of the pump beams. In the simulations, the 1550 nm beam waist is 40 μm, and the phase mismatch is $4.10\times10^{-4}$ /μm. Curves a to f correspond to the change of the 795 nm beam waist from 30 μm to 80 μm in steps of 10 μm.

Additionally, we find that the spatial structure varies asymmetrically in the positive ($\Delta k > 0$) and negative ($\Delta k < 0$) phase mismatch regions. The simulation results are shown in Figure 4. Rows P1 and P2 show the results for positive phase mismatch, while rows N1 and N2 show the corresponding profiles for negative phase mismatch with the same absolute values of phase mismatch. For the positive phase mismatch, the three phases have the same sign, and thus the spatial phase modulation inside the crystal is enhanced, while for the negative phase mismatch, the signs of the pump phases and the sign of the phase mismatch are different, and thus the spatial phase modulation is reduced. This results in a larger change period within the negative phase mismatch region than in the positive phase mismatch region with respect to the phase mismatch value. This effect can be clearly seen in Figure 4.

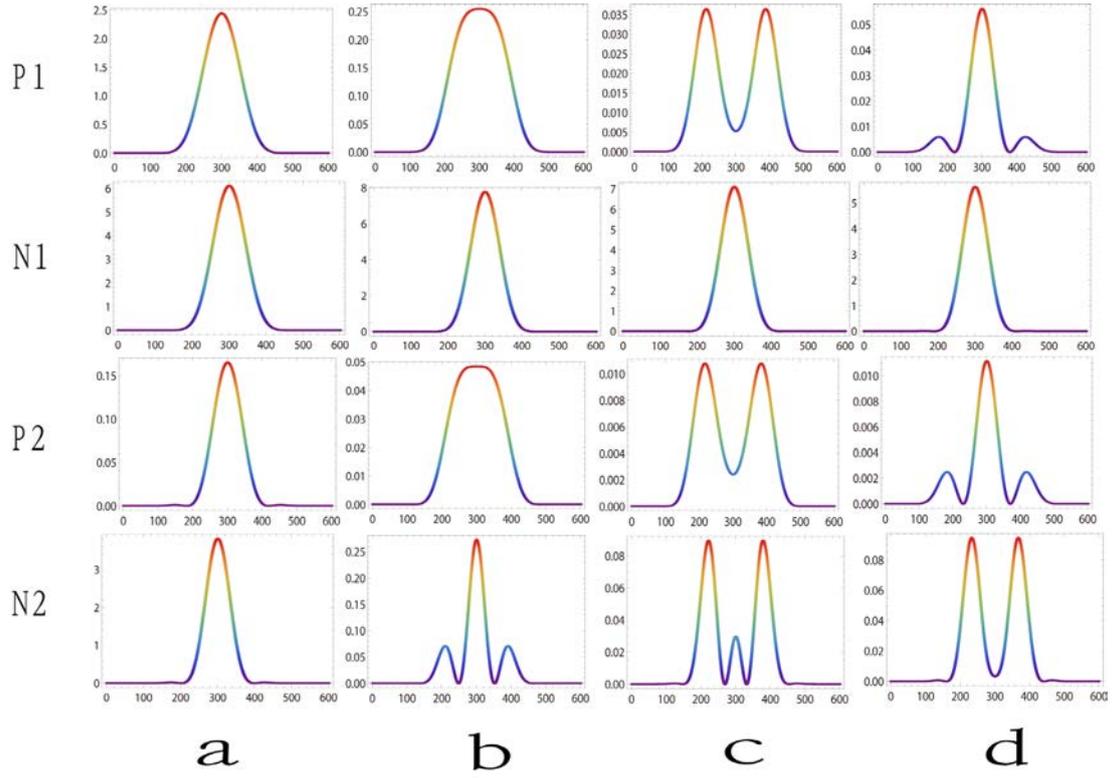

Figure 4. Simulation results of the asymmetrical SFG beam behavior in the positive and negative phase matching regions. The pump beam waists are 70 μm (795 nm) and 40 μm (1550 nm). Rows P1 and P2 are the results of positive phase mismatching, and rows N1 and N2 are the corresponding results of negative phase mismatching. The phase mismatching values in rows P1 and P2 are (1.0, 3.0, 4.0, 5.0, 6.0, 9.0, 10.0 and 11.0) × $10^{-4}$ /μm.

In conclusion, we have generated light beams with controllable spatial structures through SFG in a periodically poled crystal. Typical structures observed include the multi-ring-like BGB, the donut-like LGB, and a super-Gaussian-like beam. The spatial structure evolution with changes in the pump focusing parameters and the phase mismatch value is discussed and numerically simulated in detail. The results presented here will provide greater insight into the processes of SFG in QPM crystals. The special structures observed here may be used to generate light beams with specific spatial structures for optical trapping and manipulation applications. The changes in the spatial structures relative to the tuning of the pump wavelength can be used to construct an all-optical switch for spatial optical switching applications.

**Acknowledgements**

This work was supported by the National Fundamental Research Program of China (Grant No. 2011CBA00200), the National Natural Science Foundation of China (Grant Nos. 11174271, 61275115, 10874171), and the Innovation Fund from the Chinese Academy of Science.